# Dumb Hole Associated with an Oscillating Bubbles Cluster

Ion Simaciu[1,a], Gheorghe Dumitrescu[2], Zoltan Borsos[1,b], Viorel Drafta[3]

1 Petroleum-Gas University of Ploiești, Ploiești 100680, Romania

2 High School Toma N. Socolescu, Ploiești, Romania

3 Independent researcher

E-mail: [a] isimaciu@yahoo.com; [b] borzolh@upg-ploiesti.ro

**Abstract**

In this paper, it is shown that in interaction with an oscillating bubbles cluster the fluid becomes inhomogeneous. The radial variation of the acoustic refractive index of the fluid generates an acoustic lens with spherical symmetry. When the focal length of the lens associated with the bubbles cluster is equal to the length of the position vector then the bubbles cluster behaves like a dumb hole. Unlike the properties of a bubble, the cluster, by compression under the action of attractive internal forces, can shrink its radius to become a dumb hole.

## 1. Introduction

In a previous paper [1, 2, 3] we have shown that a packet of the spherical stationary acoustic waves and a bubble in radial oscillation acts as a convergent acoustic lens.

In this paper, we investigate the properties of another object in the acoustic world, namely an oscillating bubbles cluster. In the second section, we infer the expression of the acoustic pressure around a bubbles cluster that oscillates radially under the action of a plane wave. Some interesting properties of the refractive index around a cluster of oscillating bubbles can be found by exploring the shape obtained analogously to that in the papers [1, 2, 3].

Our derivation from the third section leads to the dependence of the refractive index on the length of the position vector in relation to the center of the bubbles cluster. When the focal length of the lens associated with the bubbles cluster is equal to the length of the position vector then the bubbles cluster behaves like a dumb hole. This phenomenon is possible because the radius of the cluster, under certain conditions, may be smaller than the acoustic radius of the cluster. Unlike the properties of a bubble [1, 2], the cluster, by compression under the action of attractive internal forces [4, 5], can shrink its radius to become a dumb hole.

This property is analogous to the property of a particles system in the electromagnetic world, such as a star, which, under the action of gravitational forces, can collapse and generate a black hole.

The fifth section is devoted to discussions and conclusions.

## 2. Acoustic pressure around an oscillating cluster

To obtain the expression of the pressure around a spherical cluster with $N$ identical bubbles ($R_{0i} = R_0$, $i = 1, 2, ..., N$) oscillating in phase, we use the expression of the pressure around an oscillating bubble inside the cluster [1, 2]

$$p'_N(r,t) \cong \frac{\rho \ddot{V}_N}{4\pi r} = \frac{\rho R_N}{r}\left(2\dot{R}_N^2 + R_N \ddot{R}_N\right) \cong \frac{\rho R_N^2 \ddot{R}_N}{r} \cong \frac{\rho R_0^3 \omega^2 x_N}{r}. \tag{1}$$

Because the oscillations of the bubbles in the cluster are coupled, the dimensionless elongation of the oscillations is [4, 5]

$$x_N = a_N \cos(\omega t + \varphi_N) \tag{2}$$

with

$$a_N = \frac{A}{\rho R_0^2 \left[\left(\omega^2(1+N_c) - \omega_0^2\right)^2 + 4\beta^2 \omega^2\right]^{1/2}}, \quad \varphi_N = \arctan \frac{2\beta \omega}{\omega^2(1+N_c) - \omega_0^2}. \tag{3}$$

Substituting Eqs. (2) and (3) in Eq. (1), results

$$p'_N(r,t) \cong \frac{R_0 \omega^2 A \cos(\omega t + \varphi_N)}{r\left[\left(\omega^2(1+N_c) - \omega_0^2\right)^2 + 4\beta^2 \omega^2\right]^{1/2}}. \tag{4}$$

The pressure produced outside the cluster, when $r \geq R_c$, is approximated by the expression

$$p'_{Ncl}(r,t) \cong \frac{NR_0 \omega^2 A \cos(\omega t + \varphi_N)}{r\left[\left(\omega^2(1+N_c) - \omega_0^2\right)^2 + 4\beta^2 \omega^2\right]^{1/2}}. \tag{5}$$

At the resonance of velocity, $\omega = \omega_{0Nr} = \omega_{0r}/\sqrt{1+N_c} < \omega_{0r}$, the amplitude and the phase (3) become:

$$a_{N,res} = \frac{A}{2\rho R_0^2 \beta_{N,res} \omega_{0N,res}}, \quad \varphi_{N,res} = \arctan \infty = \frac{\pi}{2}. \tag{6}$$

At the resonance, Eqs. (4) and (5) become:

$$p'_{N,res}(r,t) \cong \frac{R_0 \omega_{0N,res} A \sin(\omega_{0N,res} t)}{2r \beta_{N,res}}, \tag{7}$$

$$p'_{Ncl,res}(r,t) \cong \frac{NR_0 \omega_{0N,res} A \sin(\omega_{0N,res} t)}{2r \beta_{N,res}}. \tag{8}$$

If the damping coefficient is approximated by

$$\beta_{N,res} = \beta_{0Nac,res} + \beta_{0N\mu,res} \cong \beta_{0Nac,res} = \frac{\omega_{0N,res}^2 R_0}{2u}, \tag{9}$$

Eqs. (7) and (8) become

$$p'_{N,res}(r,t) \cong \frac{uA\sin(\omega_{0N,res}t)}{r\omega_{0N,res}} = \frac{uA\sqrt{1+N_c}\sin(\omega_{0N,res}t)}{r\omega_0}, \tag{10}$$

$$p'_{Ncl,res}(r,t) \cong Np'_{N,res}(r,t) = \frac{N\sqrt{1+N_c}\,uA\sin(\omega_{0Nr}t)}{r\omega_{0r}}, \quad r > R_c. \tag{11}$$

## 3. The refractive index generated by an oscillating bubbles cluster

In following we will use the expression of refractive index [1, 2] for inhomogeneous medium

$$\langle n_{acl}(p)\rangle_t \cong 1 + \frac{\zeta(3\zeta+1)}{2(2\zeta+1)^2} \frac{\langle (p'_{Ncl}(r,t))^2\rangle_t}{p_0^2}. \tag{12}$$

Substituting Eqs. (5) and (11) in Eq. (12), yields:

$$\langle n_{acl}(p)\rangle_t \cong 1 + \frac{N^2\zeta(3\zeta+1)R_0^2}{4(2\zeta+1)^2 r^2} \frac{A^2\omega^4}{p_0^2\left[\left(\omega^2(1+N_c)-\omega_{0r}^2\right)^2 + 4\beta_r^2\omega^2\right]}, \tag{13}$$

$$\langle n_{acl,res}(p)\rangle_t \cong 1 + \frac{N^2(1+N_c)\zeta(3\zeta+1)}{4(2\zeta+1)^2 r^2} \frac{A^2 u^2}{p_0^2\omega_0^2}. \tag{14}$$

The dependence of the acoustic refractive index on the length of the position vector $r$ reveals the acoustic inhomogeneity, with spherical symmetry, of the liquid around the oscillating cluster. That is the liquid surrounding the cluster behaves like a spherical lens. Given an acoustic wave which is plane, and which is deflected when it passes at the minimum distance from the center of the bubble, the focal length of the spherical lens is [6]:

$$\frac{1}{f_{cl}} = \frac{2}{r}\left(\langle n_{acl}\rangle_t - 1\right) = \frac{N^2\zeta(3\zeta+1)}{2(2\zeta+1)^2}\frac{R_0^2}{r^3}\frac{A^2}{p_0^2}\frac{\omega^4}{\left[\left(\omega^2(1+N_c)-\omega_0^2\right)^2 + 4\beta^2\omega^2\right]} \tag{15}$$

and

$$\frac{1}{f_{cl,res}} = \frac{2}{r}\left(\langle n_{acl,res}\rangle_t - 1\right) = \frac{N^2(1+N_c)\zeta(3\zeta+1)}{2(2\zeta+1)^2}\frac{A^2}{p_0^2}\frac{u^2}{\omega_0^2 r^3} \tag{16}$$

or

$$f_{cl} = \frac{2(2\zeta+1)^2}{N^2\zeta(3\zeta+1)}\frac{r^3}{R_0^2}\frac{p_0^2}{A^2}\frac{\left(\omega^2(1+N_c)-\omega_0^2\right)^2 + 4\beta^2\omega^2}{\omega^4} > 0 \tag{17}$$

and

$$f_{cl,res} = \frac{2(2\zeta+1)^2}{N^2(1+N_c)\zeta(3\zeta+1)} \frac{p_0^2}{A^2} \frac{\omega_0^2 r^3}{u^2} > 0. \tag{18}$$

Since the focal length is positive, it turns out that the spherical lens associated with the cluster is convergent.

## 4. Oscillating bubbles cluster and dumb hole

The convergent acoustic lens, associated with a cluster, behaves like a dumb hole if

$$f(r) = r. \tag{19}$$

Then, according to Eqs. (17) and (18), the corresponding dumb hole radius becomes

$$r_{acl} = R_0 \frac{N\sqrt{\zeta(3\zeta+1)}}{\sqrt{2}(2\zeta+1)} \frac{A}{p_0} \frac{\omega^2}{\sqrt{(\omega^2(1+N_c)-\omega_0^2)^2 + 4\beta^2\omega^2}} = Nr_{abcl}. \tag{20}$$

$$r_{acl,res} = \frac{N\sqrt{1+N_c}\sqrt{\zeta(3\zeta+1)}}{\sqrt{2}(2\zeta+1)} \frac{Au}{p_0\omega_0} \simeq \frac{N\sqrt{N_c}\sqrt{\zeta(3\zeta+1)}}{\sqrt{2}(2\zeta+1)} \frac{Au}{p_0\omega_0} = Nr_{abcl,res}. \tag{21}$$

In Eqs. (20) and (21), we identify the quantities: the acoustic radius of a coupled bubble $r_{abcl}$ (different from the acoustic radius of a free bubble given by relation (17) in [1, 2]) and the acoustic radius of a coupled bubble, at resonance $r_{abcl,res}$:

$$r_{abcl} = R_0 \frac{\sqrt{\zeta(3\zeta+1)}}{\sqrt{2}(2\zeta+1)} \frac{A}{p_0} \frac{\omega^2}{\sqrt{(\omega^2(1+N_c)-\omega_0^2)^2 + 4\beta^2\omega^2}}, \tag{22}$$

$$r_{abclr} = \frac{\sqrt{N_c}\sqrt{\zeta(3\zeta+1)}}{\sqrt{2}(2\zeta+1)} \frac{Au}{p_0\omega_0}. \tag{23}$$

Since, according to Eq. 25 of the paper [5], $N_c \cong N(3R_0/(2R_c)) \gg 1$, Eqs. (21) and (23) becomes

$$r_{acl,res} = \frac{N^{3/2}\sqrt{3R_0}\sqrt{\zeta(3\zeta+1)}}{2\sqrt{R_c}(2\zeta+1)} \frac{Au}{p_0\omega_0} = Nr_{abcl,res}. \tag{24}$$

and

$$r_{abcl,res} = \frac{\sqrt{3NR_0}\sqrt{\zeta(3\zeta+1)}}{2\sqrt{R_c}(2\zeta+1)} \frac{Au}{p_0\omega_0}. \tag{25}$$

Replacing the natural angular frequency, $\omega_0 = \{3\gamma[p_0/(\rho R_0^2) + 2\sigma/(\rho R_0^3)] - 2\sigma/(\rho R_0^3)\}^{1/2} = [p_{eff}/(\rho R_0^2)]^{1/2}$ [7], in Eqs. (24) and (25), results:

$$r_{acl,res} = \frac{R_0 N^{3/2}\sqrt{3R_0}\sqrt{\zeta(3\zeta+1)}}{2\sqrt{R_c}(2\zeta+1)} \frac{A}{p_0}\sqrt{\frac{\rho u^2}{p_{eff}}} = Nr_{abcl,res}, \tag{26}$$

$$r_{abcl,res} = \frac{R_0\sqrt{3NR_0}\sqrt{\zeta(3\zeta+1)}}{2\sqrt{R_c}(2\zeta+1)} \frac{A}{p_0}\sqrt{\frac{\rho u^2}{p_{eff}}}. \tag{27}$$

Unlike the free oscillating bubble which cannot become a dumb hole [1, 2], the cluster can be compressed so that for $R_c \leq r_{acl,res}$ it becomes a dumb hole. In the particular case, $R_c = r_{acl,res}$, by replacing this condition in Eqs. (26) and (27), results an expressions independent of $R_c$

$$r_{acl,res} = NR_0 \left[ \frac{3\zeta(3\zeta+1)}{4(2\zeta+1)^2} \left(\frac{A}{p_0}\right)^2 \frac{\rho u^2}{p_{eff}} \right]^{1/3} = Nr_{bacl,res}, \quad (28)$$

$$r_{bacl,res} = R_0 \left[ \frac{3\zeta(3\zeta+1)}{4(2\zeta+1)^2} \left(\frac{A}{p_0}\right)^2 \frac{\rho u^2}{p_{eff}} \right]^{1/3} = r_{ba,res} \left[ \frac{6(2\zeta+1)}{\sqrt{\zeta(3\zeta+1)}} \frac{p_0}{A} \sqrt{\frac{p_{eff}}{\rho u^2}} \right]^{1/3} > r_{ba,res}, \; p_0 \gg A, \quad (29)$$

whit the expression of acoustic radius $r_{ba,res}$, according to Eq. (17) of the paper [1, 2].

The expression of the acoustic radius of the cluster, Eq. (28) $r_{acl,res} = Nr_{abcl,res}$, is analogous to the gravitational radius of a system of $N$ particles with mass $m$ and gravitational radius $r_{gm} \simeq (Gm)/c^2$ [8]

$$r_{gM} \simeq \frac{G(Nm)}{c^2} = Nr_{gm}. \quad (30)$$

This analogy suggests that two oscillating clusters also interact through an attractive force generated as an effect of the absorption of the energy of the exciting wave. We will investigate this phenomenon in a future paper.

## 5. Discussions

In this paper, we have emphasized that around an oscillating bubbles cluster the fluid becomes inhomogeneous.

The oscillating bubbles cluster behaves like a convergent spherical lens that deviates the acoustic waves.

The cluster becomes a dumb hole when it is compressed so the cluster radius becomes equal with the acoustic radius of the cluster $R_c = r_{acl}$.

It follows that the clusters not only scatter the acoustic waves but also absorb them when the impact distance is less than or equal to the acoustic radius of the cluster $r \leq r_{aclr}$. According to the analogy with the phenomena in the electromagnetic world, the clusters interact both electro-acoustically (like two oscillating bubbles [9, 10]) and attract each other (gravito-acoustically) [11].